# Reliability of meta-analysis of an association between ambient air quality and development of asthma later in life

S. Stanley Young,[1] Kai-Chieh Cheng,[2] Jin Hua Chen,[2] Shu-Chuan Chen,[3] Warren B. Kindzierski[4]

**Affiliations**

[1]CGStat, Raleigh, NC, USA

[2]Graduate Institute of Data Science, Taipei Medical University, Taipei City, Taiwan

[3]Department of Mathematics and Statistics, Idaho State University, Pocatello, ID, USA

[4]School of Public Health, University of Alberta, Edmonton, Alberta, Canada

**Corresponding author:**

Warren B. Kindzierski, School of Public Health, University of Alberta, Edmonton, Alberta, T6G 1C9, Canada. Email: warrenk@ualberta.ca.; phone: 780-492-0382; fax: 780-492-0364.

## Abstract

Claims from observational studies often fail to replicate. A study was undertaken to assess the reliability of cohort studies used in a highly cited meta-analysis of the association between ambient nitrogen dioxide (NO2) and fine particulate matter (PM2.5) concentrations early in life and development of asthma later in life. The numbers of statistical tests possible were estimated for 19 base papers considered for the meta-analysis. A p-value plot for NO2 and PM2.5 was constructed to evaluate effect heterogeneity of p-values used from the base papers. The numbers of statistical tests possible in the base papers were large – median 13,824 (interquartile range 1,536−221,184; range 96−42M) in comparison to statistical test results presented. Statistical test results drawn from the base papers are unlikely to provide unbiased measures for meta-analysis. The p-value plot indicated that heterogeneity of the NO2 results across the base papers is consistent with a two-component mixture. First, it makes no sense to average across a mixture in meta-analysis. Second, the shape of the p-value plot for NO2 appears consistent with the possibility of analysis manipulation to obtain small p-values in several of the cohort studies. As for PM2.5, all corresponding p-values fall on a 45-degree line indicating complete randomness rather than a true association. Our interpretation of the meta-analysis is that the random p-values indicating no cause−effect associations are more plausible and that their meta-analysis will not likely replicate in the absence of bias. We conclude that claims made in the base papers used for meta-analysis are unreliable due to bias induced by multiple testing and multiple modelling (MTMM). We also show there is evidence that the heterogeneity across the base papers used for meta-analysis is more complex than simple sampling from a normal process.

# 1 Introduction

In this paper we examine what we contend is an important contributing cause of the current replication problem in science [1−5] – multiple testing and multiple modelling (MTMM) [6]. Multiple testing involves statistical null hypothesis testing of many separate predictor variables against numerous dependent variables and multiple modelling involves using multiple model selection procedures or different model forms (e.g., logistic regression, times series, case crossover, etc.). To support our contention, we build our paper around a specific research claim made in a meta-analysis that exposure to ambient air quality early in life is associated with the development of asthma later in life [7,8].

In the paper we discuss what is known about causes and contributing risk factors to development of asthma and how epidemiological observational studies are used in meta-analysis to support an quality−asthma development claim. We then introduce and use two straightforward statistical methods – analysis search space and p-value plots – to independently examine the reliability of the meta-analysis [7,8]. We believe that these methods can be used by other researchers to examine the reliability of other research claims made in meta-analysis in the biomedical sciences field and as an aid to judge whether a meta-analysis will replicate.

## 1.1 Replication problem in science

The US Centers for Disease Control and Prevention (CDC), US Environmental Protection Agency (EPA) and the World Health Organization (WHO) all indicate that poor air quality conditions – for example, high levels of ozone and fine particulate matter (PM2.5) – may make asthma symptoms worse and trigger asthma attacks in people. For example:

- The US CDC [9] indicates that outdoor air quality can trigger asthma attacks.

- The US EPA Integrated Science Assessment for Particulate Matter [10] indicates that recent epidemiologic studies continue to show associations between short-term PM2.5 exposure and asthma exacerbation (asthma attack). In addition, although more limited epidemiologic studies exist, US EPA [10] indicates that evidence from these studies show associations between long-term PM2.5 exposure and asthma development in children and asthma prevalence in children.
- The WHO [11] also indicates that both short- and long-term exposure to ambient air quality can lead to reduced lung function, respiratory infections and aggravated asthma in children and adults.

The US CDC and EPA and WHO statements rely heavily on epidemiologic observational studies. Observational studies involve investigators observing individuals without manipulation or intervention. However, it has been shown that research claims coming from observational studies [12] and even experimental studies [13] often fail to replicate. Scientific opinions vary as to how serious the replication problem is, but Baker [14] reported that about 90% of 1,576 scientists surveyed by *Nature* consider the replication problem to be a severe crisis (52%) or at least a slight crisis (38%).

Elsewhere there is agreement that a replication problem exists in scientific research as three published books [1−3] and reports from the National Association of Scholars [4] and the National Academies of Sciences, Engineering, and Medicine [5] discuss the problem and what might be done about it. In short, research claims made in observational studies can be unreliable and fail to replicate. Even in situations where a claim replicates, it does not guarantee that the claim is reliable [5]. Franco et al. [15] and Nissen et al. [16] point out that negative studies are often not reported by authors; and even if they are submitted, editors often reject them out of hand, so a false positive claim can mistakenly be presumed as established fact.

Randall and Wesler [4] report that improper research techniques, lack of accountability, disciplinary and political groupthink, and a scientific culture biased toward producing positive results have produced a critical state of affairs in a wide range of scientific and social-scientific disciplines, from epidemiology to social psychology. They also report that 'uncontrolled researcher freedom' makes it easy for researchers to err in many different ways, including [4]:

- Having a poor understanding of statistical methodology.
- Consciously or unconsciously using bias during data manipulation to produce desired outcomes.
- Choosing between multiple measures of a variable and often deciding to use those that provide statistically significant results (and ignore those that do not provide statistically significant results).
- Using illegitimate manipulations of research techniques.

The National Academies of Sciences, Engineering, and Medicine [5] acknowledges that statistical inference plays an oversized role in science replicability discussions due to frequent misuse of statistics such as the p-value and its threshold for determining statistical significance. P-values – particularly small p-values – do not provide a direct estimate of how likely a result is true or of how likely the null hypothesis ('there is no effect') is true [17]. Moreover, small p-values do not convey whether a result is clinically or biologically significant, or large enough to have practical value. In reality, p-values are dependent on the data, statistical method used, assumptions made and the appropriateness of the assumptions [17]. The National Academies of Sciences, Engineering, and Medicine [5] further identify that inappropriate reliance on statistical significance may lead to biases in research reporting and publication.

### 1.2 Meta-analysis

A meta-analysis offers a window into the reliability of a research claim. A meta-analysis examines a claim by taking a summary statistic along with a measure of its reliability from multiple studies found in the literature [18]. The statistics are combined to give what in theory is supposed to be a better [19] or more reliable [20] estimate of an effect. Two key assumptions of a meta-analysis are that the estimates coming from base papers into the analysis are an unbiased estimate of the effect of interest [21] and that meta-analysis of multiple studies offers a pooled estimate with increased precision [22], i.e., there is one etiology in operation.

Young and Kindzierski [23] counter that – as researchers can often ask a lot of questions and compute many models in an observational study – any statistics coming from such a study cannot be guaranteed to be unbiased. For example, a result from an observational study (e.g., a treatment effect) associated with a small p-value may itself be an order statistic, the largest treatment difference among many examined. A demonstration of this is nutritional food frequency studies, where the amounts (e.g., servings per day) of over 100 different types of foods consumed by study subjects are recorded, and after some follow-up health symptoms are recorded and compared against food quantities consumed [24].

Such studies may have many hundreds of possible questions at issue due to the multiple foods and multiple health symptoms recorded. A treatment difference from such a study might well be extreme, due to chance and certainly should not be taken as an unbiased estimate due to the many hundreds of possible questions at issue. Also, modelling is typically used to reduce variability and aggressive modelling may lead to an underestimate of variability. Treatment differences might be biased high due to chance and estimates of variability might be biased low due to multiple modelling.

It is common in a meta-analysis to examine the consistency of a result across multiple combined studies. For example, any kind of variability across studies in a meta-analysis may be termed ‘heterogeneity’ [25]. Heterogeneity arises because of effects in the subjects which the studies represent are not the same. Heterogeneity is examined by looking at the ‘across study’ variability versus the ‘within study’ variability [26,27]. Very often there is more heterogeneity in meta-analysis than one would expect by chance. If one assumes that summary statistics are coming from a ‘consistent process’ with extra variability [28], there is a meta-analysis process to give a combined (weighted) estimate of an effect. However, the heterogeneity might not be as simple as selecting values from a normal distribution with larger variability. This will be explored further in our paper.

### 1.3 Causes and risk factors for development of asthma

Asthma is associated with three principal characteristics [29,30]: (1) variable airways obstruction, (2) airway hyperresponsiveness, and spasm with wheezing and coughing and (3) airway inflammation. Busse and Lemanske [31] characterize asthma clinically by episodic, reversible obstructive airways obstruction that variably presents as a variety of symptoms from cough to wheezing, shortness of breath, or chest tightness.

The diagnosis of asthma/reactive airways is challenging; for example, if an adult who has smoked for a time develops bronchitis they will demonstrate reactive airways that are triggered by various mechanisms – e.g., fumes, cold air, exercise – and so they start to look like late onset asthmatics. Whereas asthma that starts young does not start as babies – they develop their asthma when they are toddlers. Asthmatics that develop a serious problem in their preadolescent years could easily be hyperresponders who would have benefited from desensitization from lack of immune challenges in postneonatal immune system development.

The underlying pathology of asthma, regardless of its severity, is chronic inflammation of the airways and reactivity/spasm of the airways. A combined contribution of genetic predisposition and non-genetic factors (Figure 1) account for divergence of the immune system towards T helper (Th) type 2 cell responses that include production of pro-inflammatory cytokines, Immunoglobulin E (IgE) antibodies and eosinophil infiltrates (circulating granulocytes) known to associate with asthma [30]. The release of pro-inflammatory cytokines that cause airway narrowing is responsible for cough, shortness of breath, wheezing and chest tightness characteristic of the asthmatic state [32]. But this fails to account for beta stimulus of bronchiolar muscles that increases airway spasm. Airway inflammation causes secretions and contributes to edema (swelling) but is does not cause hyperresponsiveness.

There is ongoing debate as to whether asthma is one disease or several different diseases that include airway inflammation; however two thirds (or more) of asthmatic patients have an allergic component to their disease and are felt to have allergic asthma [33]. Not enough is currently known to rule out allergic causes for a vast majority of asthma problems. As for development of asthma [34], the disease frequently first expresses itself early in the first few years of life arising from a combination of host (genetic) and non-genetic factors (Table 1). Most investigators would agree there is a major hereditary contribution to the underlying causes of asthma and allergic diseases [35].

### 1.4 Objective of study

Given acknowledged lack of replication in scientific studies today [1−3] and recent attention to this issue by prominent organizations such as the National Association of Scholars [4] and the National Academies of Sciences, Engineering, and Medicine [5], we undertook an evaluation of a highly cited meta-analysis of the association between inferred exposure to ambient air quality early in life and development of asthma later in life by Anderson et al. [7,8]. As of September 7, 2020, this study had 206 Google Scholar citations.We employed statistical methods that are non-traditional to mainstream environmental epidemiology in our evaluation– analysis search space and p-value plots – to examine two specific properties of the meta-analysis:

1. Whether claims made in the base papers used for meta-analysis by Anderson et al. are potentially unreliable due to bias induced by multiple testing and multiple modelling, MTMM [6,23].
2. Whether there is evidence that the heterogeneity in data across base papers used for the meta-analysis is more complex than simple sampling from a single normal process [28].

These statistical methods can be used to independently examine the reliability of a meta-analysis making a research claim. We believe that these methods can be used as an aid to judge whether a meta-analysis will replicate.

## 2 Methods

### 2.1 US national trends in asthma prevalence and air quality

We initially wanted to understand asthma prevalence in the US population in relation to ambient air quality. To do this, we accessed asthma prevalence data from the US Centers for Disease Control and Prevention (Atlanta, GA) [42,43] for the time period 1990−2017. The prevalence data [42,43] are from annual national surveys conducted by National Center for Health Statistics (NCHS), US Department of Health & Human Services and are self-reported by respondents to the National Health Interview Survey.

We also wanted to understand changes in air pollutant concentrations in the US over a similar period. To do this we accessed US annual national air pollutant concentration averages over the same time period (1990−2017) from US Environmental Protection Agency [44]. Weatherhead et al. [45] note that meaningful – i.e., statistically significant – changes over time, or trends, for datasets that are limited in time duration (e.g., only a couple of years) are unlikely to be truly representative of the trends that are actually occurring; and that this would only be captured with a much longer time period of two to three decades. A 27−year period used here is sufficiently long enough to depict meaningful changes over time for air quality components in the US.

### 2.2 Background on the Anderson et al. meta-analysis

It is generally accepted that there are two classes of causes of asthma [34] – primary and secondary. 'Primary' causes relate to the increase in risk of developing the disorder (e.g., asthma). Whereas 'secondary' causes relate to precipitation of asthma attacks (exacerbations). The Anderson et al. meta-analysis [7,8] focused on cohort studies of the association between ambient air quality components and development of asthma later in life, and hence on the primary causes of asthma.

We initially filed a public standard operating procedure (SOP) for our evaluation methods with the Center for Open Science 'open science framework' [46]. Anderson et al. [7,8] conducted a systematic review and meta-analysis of cohort studies of the association between two air quality components – particulate matter with aerodynamic equivalent diameter ≤2.5 micron (PM2.5) and nitrogen dioxide (NO2) – and incidence of asthma. To increase the number of estimates for each air quality parameter–outcome pair, Anderson et al. scaled results for studies of particulate matter with aerodynamic diameter <10 μm (PM10) to PM2.5 using a factor of 0.65 and of oxides of nitrogen (NOx) to nitrogen dioxide (NO2) using a factor of 0.44.

Incidence was defined as the incidence of diagnosed asthma or of new wheeze symptom between two assessments or, in birth cohorts followed up to 10 years of age, a lifetime prevalence estimate of asthma or wheeze symptom. Anderson et al. indicated that most cohort studies they used for their meta-analysis reported outcomes as odds ratios (ORs), but some reported them as relative risks (RRs) or hazard ratios (HRs). Anderson et al. indicated that all three quantitative health outcome estimates were combined for their meta-analysis as they indicated that the outcome of interest (asthma) is quite common but the effect size is relatively small. The quantitative health outcome estimates (referred to as effect estimates or EEs here) and their 95% confidence intervals (CIs) of the air quality−asthma incidence association were standardized to a 10 $\mu g/m^3$ increment of each air quality component.

Anderson et al. identified 17 cohorts studies in their review – eight birth cohorts and nine child/adult cohorts – of relationships between outdoor air quality and the incidence of asthma or wheeze symptom with a total of 99 EEs from 24 published articles. Most cohorts were based on inferred within community exposure contrasts dominated by traffic pollution. Twelve of the cohorts reported at least one positive statistically significant association ($p$<.05) between an air quality component and a measure of incidence. Of the total of 99 EEs, only 29 were positive and statistically significant (i.e., $p$<.05) and the remaining 70 were negative.

Thirteen of their cohorts reported results for oxides of nitrogen (NOx), mostly as nitrogen dioxide (NO2), and were used for their meta-analysis of NO2. Of the 13 cohorts used, two had multiple publications. Anderson et al. did not specify which publications they drew upon for their EEs and CIs of these two cohorts. Also, five cohorts were used by for their meta-analysis of PM2.5. Of the five cohorts used, four had multiple publications. Again, Anderson et al. did not specify which publications they drew upon for their EEs and CIs of these four cohorts.

It is important to point out that published epidemiologic studies with negative (non-statistically significant) results are more likely to remain unpublished than studies with positive results in scientific literature [17]. This results in a distortion of effects in epidemiological literature and any systematic review or meta-analysis of these studies may be biased [47,48] because they are summarizing information and data from a misleading, selected body of evidence [5,49].

Anderson et al. reported the following results of their meta-analysis: (i) for 13 cohort studies with NO2 estimates, the random EE was 1.15 (95% CI 1.06 to 1.26) per 10 $\mu g/m^3$, and (ii) for five cohort studies with estimates for PM2.5, the random EE was 1.16 (95% CI 0.98 to 1.37) per 10 $\mu g/m^3$. Finally, Anderson et al. state in their Abstract that… "*The results are consistent with an effect of outdoor air pollution on asthma incidence*."

## 2.2 Analysis search space

We refer to the published studies used in the Anderson et al. [7,8] meta-analysis as 'base papers'. Analysis search space (or search space counts) represents an estimate of the number of statistical tests, in this case, of exposure-disease combinations tested in a base paper. Our interest in analysis search space is explained further. There is flexibility available to researchers to undertake a range of statistical tests and use different statistical models during an observational study before selecting, using and reporting only a portion of the test and model results [23]. This researcher flexibility is commonly referred to as ‘researcher degrees of freedom’ in the psychological sciences by Wicherts et al. [50]. Base papers with large search space counts suggest the use of a large number of statistical tests and statistical models and the potential for researchers to search through and only report only a portion of their results (i.e., positive, statistically significant results).

We started with the Anderson et al. 24 base papers. We identified the 13 cohort studies with NO2 EEs and five cohort studies with EEs for PM2.5 from these articles. We then separately estimated the analysis search space (number of statistical tests that may have been conducted) in 19 of 24, or 80%, of the base papers. Electronic copies of the 19 selected bases papers (and any corresponding electronic supplementary information files) were obtained and read. Refer to Supplemental Information (SI) 1 for a listing of the 19 base papers that we used. We then separately counted the number of outcomes, predictors, time lags and covariates reported for each cohort study. Covariates can be vague as they might be mentioned anywhere. Specifically, analysis search space of a cohort study was estimated as follows:

- The product of outcomes, predictors, and time lags = number of questions at issue (Questions = outcomes x predictors x lags).
- A covariate may or may not act as a confounder to a predictor variable and the only way to test for this is to include/exclude the covariate from a model. As it can be in or out of a model, one way to approximate the ‘modelling options’ is to raise 2 to the power of the number of covariates (Models = $2^k$, where $k$ = number of covariates).
- Questions x Models = an approximation of analysis search space (Search Space).

Three examples of how analysis search space is estimated in observational (cohort) studies is provided in SI 2. Young and Kindzierski [23] indicate that estimates of analysis search space are considered to be lower bound approximations and, what is presented here is based on information that is reported in each cohort study base paper that we evaluated.

### 2.3 P-value plots

P-value plots following the ideas of Schweder and Spjøtvoll [51] were developed to check the distribution of the set of p-values from the EEs and CIs presented by Anderson et al. [7,8] – 13 for NO2 and five for PM2.5. The p-value can be defined as the probability, if nothing is going on, of obtaining a result equal to or more extreme than what was observed. The p-value is a random variable derived from a distribution of the test statistic used to analyze data and to test a null hypothesis.

Hung et al. [52] indicate that under the null hypothesis, the p-value is distributed uniformly over the interval 0 to 1 regardless of sample size. A distribution of true null hypothesis points in a p-value plot should form a straight line [51]. A plot of p-values sorted by rank corresponding to true null hypothesis points should conform to a near 45 degree line. The plot can be used to assess the validity of a false claim being taken as true and, specific to our interest, can be used to examine the claims made in the base papers used for meta-analysis.

A p-value plot was constructed for p-values of the 18 estimates – 13 for NO2 and five for PM2.5 – and interpreted as follows after Schweder and Spjøtvoll [51]:

- P-values were first computed from the EEs and CIs (assuming a symmetric CI) using JMP statistical software (SAS Institute, Cary, NC) and the resulting p-values were ordered from smallest to largest and plotted against the integers, 1, 2, 3, …

- If p-value results are random (i.e., a true null relationship), the p-value plot should roughly follow a 45-degree line indicating a uniform distribution.
- Alternatively if a true relationship exists, p-value results should all be on a roughly straight line with a slope considerably less than 45 degrees [23].
- If the plotted points exhibit a bilinear shape, then the p-values used for meta-analysis comprise a mixture and a general (over-all) claim is not supported. In addition, the p-value reported for the overall claim in the meta-analysis is not valid.

To assist in interpretation of the behavior of p-value plots for the Anderson et al. meta-analysis data, we also searched out the literature and constructed and show p-value plots for 'plausible true null' and 'plausible true alternative' hypothesis outcomes based on meta-analysis of observational datasets. Hung et al. [52] note the distribution of the p-value under the alternative hypothesis – where the p-value is a measure of evidence against the null hypothesis – is a function of both sample size and the true value or range of true values of the tested parameter. These p-value plots are presented and further explained in SI 3.

## 3 Results

### 3.1 US national trends in asthma prevalence and air quality

Data from the National Health Interview Survey (NHIS) was used to present prevalence of asthma in the US. NHIS is a complex, multi-stage, probability sample survey conducted annually by the US Bureau of the Census for the National Center for Health Statistics [56]. Information is collected during in-home interviews of the civilian noninstitutionalized US population on a variety of health issues. Survey weights are used to calculate population-based estimates from NHIS respondents. Estimates of asthma prevalence indicate the percentage of the population with asthma at a given point in time (i.e., at the time of the health survey).

Figure 2 shows population-weighted annual prevalence of asthma in the US for a 27−year period 1990−2017 as reported by National Center for Health Statistics, National Health Interview Surveys [42,43]. Asthma prevalence in the US population increased from 4.2% in 1990 to 7.9% in 2017; although it has been relatively stable at ~8% since 2006. No data were available for the years 1997−2000.

Table 2 summarizes the percent change in US annual national air pollutant concentration averages [44] and annual prevalence of asthma in the US (population-weighted) [42,43] over the 1990−2017 period. As shown in Table 2, all air quality components of interest to the US Environmental Protection Agency declined (range 22−88%) over the 27−year period. In particular, US ambient NO2 (PM2.5) concentrations declined by 50% (40%); on the other hand, prevalence of asthma in the US (population-weighted) increased by 88% during the same period. These conflicting trends suggest that other factors (refer to Figure 1 and Table 1), rather than air quality components, may be more important in the development of asthma later in life.

### 3.2 Analysis search space

Estimated analysis search spaces for the 19 based papers representing 14 cohort studies from Anderson et al. [7,8] are presented in Table 3. From Table 3, investigating multiple – 2 or more – asthma outcomes (i.e., Outcomes listed in Table 3) in the cohort studies were as common as single outcome investigations. In addition, use of multiple Predictors and Lags was very common, and so was adjusting for multiple possible Covariate confounders. While these multiple factors (i.e., outcomes, predictors, lags and covariates) seemingly represent realistic attempts to simulate/model possible exposure−disease combinations, these combinations can easily inflate the overall number of possible statistical tests performed in a single study.

Summary statistics of the possible numbers of tests in the 19 base papers representing 14 cohort studies are presented in Table 4. The median number (interquartile range, IQR) of Questions and Models was 24 (IQR 15−84) and 256 (IQR 96–3,072), respectively. The median number (IQR) of possible statistical tests (Search space) of the 19 base papers was 13,824 (IQR 1,536−221,184). Given the large numbers of possible statistical tests, results taken from the cohort studies are unlikely to offer unbiased measures for meta-analysis. Although not shown, covariates in each of the cohort studies vary considerably from study to study (the reader is referred to the original Anderson et al. supplemental files). For comparison purposes, we note that search space counts of air quality component−heart attack observational studies in published literature are similarly large – i.e., median (IQR) = 6,784 (2,600−94,208), n=14 [57], and = 12,288 (2,496−58,368), n=34 [23].

### 3.3 P-value plots

Table 5 presents EEs, CIs and calculated p-values for the 18 cohort studies. A p-value plot of the sorted p-values versus the integers is given in Figure 3. Both p-values for NO2 (indicated by solid circles, •) and PM2.5 (indicted by open circles, ο) are given in Figure 3. The plot presents as bi-linear with six p-values near or below nominal significance (.05) and the remaining p-values are >.05 and fall on an approximately 45-degree line. This is different from p-value plot behavior of both plausible true null and true alternative hypothesis outcomes (refer to Figures SI3−1 and SI3−2).

As p-value plots are standard technology, the two-component mixture of p-values in Figure 3 is a combination of studies suggesting an association and no association. However, both outcomes cannot be true. This two-component mixture appears consistent with the possibility of analysis manipulation to obtain small p-values in several of the cohort studies. A p-value (mixture) relationship does not support a general claim that inferred exposure to ambient NOx and PM2.5 early in life is associated with development of asthma later in life.

Higgens and Green [25] assert that heterogeneity will always exist in meta-analysis whether or not one can detect it using a statistical test. Statistical heterogeneity ($I^2$) quantifies the proportion of the variation in point estimates due to among−study differences. $I^2$ is a standard measure for heterogeneity and Anderson et al. [7,8] reported an $I^2$=64.1% (p<.001) for NO2 based on 13 study cohorts and $I^2$=7.4% (p=.364) for PM2.5 based on 5 study cohorts.

## 4 Discussion

We used two statistical methods – analysis search space and p-value plots – to independently examine the reliability of the Anderson et al. meta-analysis [7,8]. Our estimation of analysis search space of the base studies (Tables 3 and 4) indicated that there were large numbers of possible statistical tests performed in the base study and the results taken from these studies are unlikely to offer unbiased measures for meta-analysis. The p-value plot that we constructed for values taken from the base studies (Figure 3) shows a two-component mixture; which is inconsistent with p-value plot behavior of either plausible true null and true alternative hypothesis outcomes (Figures SI3−1 and SI3−2. Taken together, this evidence does not support the Anderson et al. meta-analysis as being a reliable investigation for other researchers to depend upon. We discuss this further below in more detail.

### 4.1 Development of asthma

From an exposure point-of-view, isolating the role of a particular factor in the development of a disease like asthma is difficult given the multi-factorial nature of causation (Figure 1), unless the factor overwhelms. A hypothesis is that outdoor air quality conditions may contribute to symptoms of asthma; although these effects are not as pronounced as those of viruses and aeroallergens [58]. However as asthma is a complex interaction between the inhaled environment and the formed elements of the airways [59], it is challenging to evaluate the role of air quality components on the prevalence of asthma in general.

For example, enhanced Immunoglobulin E antibody-mediated response to aeroallergens and resulting enhanced airway inflammation could account for increasing frequency of allergic respiratory allergy and bronchial asthma often attributed to air quality components in observational studies [60]. By examining NO2 and PM2.5 air quality component−asthma cohort observational studies, our study helps judge if this hypothesis is supported.

Today there is some new thinking as to how asthma may develop. The usual thinking on cause—effect is that if you see result B, you should look for A as causative agent. For example, Polio is caused by the polio virus. While for asthma, the new thinking is that absence (of exposure) may be a cause. As we have shown here, prevalence of asthma has been increasing over time (Figure 2) and Haahtela [61] and Dunn [62] note that asthma is more absent in rural areas and increasing more in urban than rural areas. Geographically, asthma prevalence appears to show uneven distributions related to possible differential susceptibility or resistance to asthma and allergies in, for example, rural/farm environments versus urban environments [63,64].

Although the basis of this difference is not known, the distribution pattern is presumed to have an inverse relationship to infection and is key to the ‘hygiene hypothesis’ [65,66].

The hygiene hypothesis is a notion of the etiology of asthma and atopic disorders (disorders characterized by a tendency to be hyperallergic) based on differences of western−versus−developing country and rural−versus−urban distributions of disease. This notion posits that clean environmental settings in western countries, compared to developing countries, play a role in the increase of prevalence of these disorders in the western countries [65]. In addition, it is common practice in urban populations of western countries to protect children from bacteria and microorganisms through isolation indoors and through overuse of antibacterial soaps. This practice may be harmful in not allowing robust immune challenge in postneonatal immune system development.

As examples, low levels of asthma and allergies are found with early exposure to pets [67,68], being raised on a farm [69], larger family size [70,71], attending day-care [71] and order of birth [72,73]. Lynch et al. [74] reported that in inner-city environments, children with the highest exposure to specific allergens and bacteria during their first year were least likely to have recurrent wheeze and allergic sensitization. These findings suggest that concomitant exposure to high levels of certain allergens and bacteria in early life might be beneficial.

The risk for development of asthma and atopic conditions (i.e., atopic dermatitis, allergic rhinitis (hay fever), allergic asthma) later in life may be due to a lack of early immune challenge of the postneonatal immune system by microbial or parasitic infections possibly including environmental saprophytes and gut commensal microbiota, relative to the developing innate immune system [75]. Alteration of the diversity of the immune system early in development may lead to the establishment of immune hypersensitivity ultimately leading to inflammatory pathology.

### 4.2 Interpretation of Anderson et al. meta-analysis

We take the overall approach and EEs and CIs in the Anderson et al. [7,8] meta-analysis at face value and make our interpretations from there. It is worthwhile to note that the time gap between a subject exposure and recording of an asthma effect across the cohort studies used by Anderson et al. ranged from 3 to 23 years. To our knowledge there is no stated etiology whereby a gap between exposure and an asthmatic effect could be this long (i.e., up to 23 years).

As to an air quality−asthma development hypothesis, there may be other reasons for statistical associations between ambient air quality levels in early life and development of asthma in later life in observational studies besides a direct causal relationship. Some of the reasons may relate to observational study methodology and include [1−3,23,49,76−79]:

- Improper selection of datasets for analysis.
- Improper selection of statistical models.
- Flexible choices in methods to compute statistical results, including undertaking multiple testing and multiple modelling (MTMM) without statistical correction.
- Inadequate treatment of confounders and other latent variables.
- Selective reporting of results.

Young and Kindzierski [23] note there are many aspects of choice involved in modelling air quality−health effects in observational studies. Some of these choices involve which parameters and confounding variables to include in a model, what type of lag structure for covariates to use, which interactions need to be considered, and how to model nonlinear trends [76]. Because of a large number of potential parameters and confounders that may be included in a study, some aspect of model selection is often used. Even if models are selected in an unbiased manner, different model selection strategies may lead to very different models and outcomes for the same set of data. On the other hand, inherent bias may lead researchers to choose models that provide selective outcomes to fit a story [23].

The p-value (mixture) relationship in a standard p-value plot (Figure 3) does not support a general air quality−asthma incidence claim and their specific statement in the Abstract "*The results are consistent with an effect of outdoor air pollution* [NO2 and PM2.5] *on asthma incidence*." Although Anderson et al. follow a typical statistical approach for meta-analysis, we suggest the approach may not be meaningful if EEs & CIs drawn from base studies are not unbiased and/or if the studies as a whole form a two-component mixture. We question whether the Anderson et al. results offer strong enough statistical evidence to support their claim.

Anderson et al. can be considered a standard meta-analysis… a question is selected, a computer search for relevant published papers is undertaken, papers are identified, filtered, and a final set of base papers is selected. The thesis they examined is… *whether ambient air quality early in life leads to development of asthma later in life*. Each cohort study (i.e., base paper they selected) looked at air quality, including one or more of the following air components – carbon monoxide, nitrogen dioxide, sulfur dioxide, ozone, or particulate matter (PM); however each cohort study varied in terms of the specific air components studied. They identified 18 cohorts with a total of 99 EEs that examined air quality and asthma, but they only ended up doing a formal meta-analysis on NOx (NO or NO2) and PMx (PM10 and/or PM2.5). In their meta-analysis, three outcome estimates (ORs, RRs or HRs) with upper and lower confidence limits were extracted from the base papers and a random effects analysis was computed after DerSimonian and Laird [28].

As mentioned previously, the air quality−asthma linkage is a position held by a number of prominent regulatory agencies – US CDC [9], US EPA [10] and WHO [11]. The Anderson et al. initial computer search identified 4,165 possibly relevant papers. From this, 266 papers were examined in detail and 13 cohorts studies were selected that reported on NOx and five cohort studies were selected that report on PM. A numerical meta-analysis was computed on the two datasets, NOx and PMx, separately and we computed p-values for these datasets and combined the p-values into one figure (Figure 3).

In their search, Anderson et al. also identified asthma-related effect studies for other air quality components – e.g., carbon monoxide, ozone and sulfur dioxide – yet they only chose to report on asthma-related effects for NO2 and PM. They did not explain their reasons for disregarding carbon monoxide, ozone and sulfur dioxide in their meta-analysis. As for NO2 and PM, search space counts – number of possible statistical tests conducted – in the selected base studies (Table 3 and 4) are considered large – median search space 13,824 (IQR 1,536−221,184). Given such large search spaces, there is little assurance that the numbers drawn from the base papers into their meta-analysis are unbiased. In addition, p-hacking cannot be ruled out as an explanation for small p-values reported in the base studies.

### 4.3 Heterogeneity

Regarding heterogeneity, Higgens et al. [80] assign low, moderate and high $I^2$ values of 25%, 50%, and 75% for meta-analysis. Higgens and Green [25] provide another guide to interpretation: 0−40% might not be important, 30−60% may be moderate heterogeneity, 50−90% may represent substantial heterogeneity and 75−100% represents considerable heterogeneity.

The Higgens and Green [25] and Higgens et al. [80] criteria suggest that the Anderson et al. meta-analysis of 13 NO2 study cohorts is associated with moderate to substantial heterogeneity. In particular, we note that heterogeneity in the Anderson et al. dataset is not as simple as due to an increase in across-study variability – but is a much more problematic two-component mixture (see Figure 3). Specifically, some NO2 studies have very small p-values that may suggest real causal relationships or p-hacking, whereas other p-values fall on a 45-degree line indicating complete randomness (i.e., no relationship at all). As for their meta-analysis of 5 PM2.5 study cohorts, all corresponding p-values fall on a 45-degree line indicating complete randomness (Figure 3).

### 4.4 Real versus random associations

The p-value plot of the 18 EEs/CLs used by Anderson et al. exhibits a bi-linear appearance (Figure 3) and is clearly different from p-value plots of both plausible true null and true alternative hypothesis outcomes (refer to Figures SI3−1 and SI3−2). Six p-values are below or near .05, a value often taken as 'statistically significant', and twelve p-values appear completely random – a two-component mixture. Firstly, any sort of statistical averaging – weighted or not – does not make sense for a mixture of this type. Secondly, both findings cannot be true. The evidence for real versus random associations can be examined at a deeper level (refer to Figure 4).

*Small p-values true & large p-values false* – First, suppose the small p-values represent true associations, i.e., there is a real air quality−asthma association. In our case (Figure 3), there are two small p-values – .00328 and .00732. P-values this small are often interpreted by researchers as being real. These two p-values are close to a .005 action level proposed by Johnson [81], but larger than a .001 action level proposed by Boos and Stefanski [21] for making such an interpretation. Here, the term ‘action level’ means that if the study is replicated, the replication will give a p-value less than .05.

These rules of thumb – the traditional .05 and recently proposed .005 and .001 p-value decision criteria – all presume only one statistical test and one p-value result, i.e., no multiple testing or multiple modelling (MTMM) issues, and that the result is from a well-conducted study (i.e., with randomization, blinding and blocking), which is not true for the Anderson et al. meta-analysis and its base studies. The median number of possible p-values over the base studies is 13,824. In the meta-analysis there are 18 p-values at issue, so a small p-value could easily arise by chance given this many tests [82]. When researchers have many different hypotheses, and carry out many statistical tests on the same set of data, they run the risk of concluding that there are real differences or real associations when in fact there are none [83].

Yet there is an abundance of published literature suggesting an overall statistical air quality−adverse health relationship. Specifically, there are a very large number of papers that report a statistical association between some air quality variable and an adverse health effect. For example, a Google Scholar title search of “air pollution” and “mortality” over the years 2000−2019 returned 1,550 hits (search done 11 December 2019).

If small p-values are true, one needs to have plausible explanations for large p-values in Figure 3 being false. Here, one can speculate it is possible that some of the papers have a large p-value due to poor data, methods, small sample size or just chance. However, seven of 13 NO2 p-values and all five PM2.5 p-values are greater than the traditional .05 decision criteria in Figure

3. This requires a fair bit of rationalization given the presumed careful procedure used by Anderson et al. to screen and select their study cohorts for meta-analysis.

*Small p-values false & large p-values true* – On the other side of the coin is the argument that the small p-values are false. How might this be the case? In the presence of large numbers of statistical test performed in a base study, we offer two possible ways in which a meta-analysis can fail:

- P-hacking in the base studies [84]. P-hacking is multiple testing and multiple modelling without any statistical correction [1−3]. Simple search space counts of the base studies supports p-hacking.
- Not properly controlling for covariates such that controlling for them may make the small p-values disappear.

There is a tendency among epidemiology researchers to highlight statistically significant findings and to avoid highlighting nonsignificant findings in their research papers. This behavior may be a problem, because many of these significant findings could in future turn out to be 'false positives' [83]. The epidemiological literature may be distorted as a result and any systematic review or meta-analysis of these studies may be biased [47,48] because they are summarizing information and data from a misleading, selected body of evidence [5,49].

It makes sense to step back from a detailed consideration of the Anderson et al. study. A researcher is expected to make a good case for their point of view. For example, Greven et al. [85] recently examined the Medicare Cohort Air Pollution Study (MCAPS) dataset, which included individual−level information on time of death and age on a population of 18.2 million for the period 2000–2006 with an interest in understanding a 'long-term particulate air quality exposure−acute mortality' association claim. Greven et al. [85] suggested two ways to make a good case for a positive air quality−health effect association in population-based studies is to test for ‘within location’ or ‘across location’ effects. Greven et al. [85] noted that positive ‘across location’ effects might be due to confounding whereas positive ‘within location’ effects would be less likely biased by confounding. They detected no ‘within location’ effects nor were they able to demonstrate a positive air quality−health effect association in their analysis.

As another example, Young et al. [86] recently examined air quality−acute death associations for the eight most populous California air basins, with over 2,000,000 deaths and over 37,000 exposure days over a 13-year period. Within each air basin (i.e., ‘within location’ analysis), Young et al. [86] observed no effects similar to Greven et al. [85]. Here one has to keep in mind that as sample size goes to infinity, the standard error goes to zero, so any small ‘across location’ effect has a good chance of being due to bias.

The bias issue can largely be controlled using a new method – Local Control [87]. Its basis is simplicity itself. One clusters the objects into many small clusters and does an analysis within each cluster. One can then observe how the analysis result changes (or not) across clusters. Obenchain and Young [87] applied Local Control to a historical air quality (total suspended particulate)−mortality dataset describing a ‘natural experiment’ initiated by the federal Clean Air Act Amendments of 1970 (specifically, the Chay et al. dataset [88]). Obenchain and Young [87] confirmed the Chay et al. [88] result of no total suspended particulate−mortality association. Thus the control of confounding is important. Variables can be put into a model or an analysis and can be restricted to limited geographic regions thereby reducing the influence of confounding factors.

## 5 Conclusions

Prevalence of asthma in the US (population-weighted) increased by 88% over the 27−year period 1990−2017; while US ambient NO2 (PM2.5) concentrations declined by 40% (50%) during the same period. These conflicting trends suggest that factors other than ambient NO2 and PM2.5 concentrations early in life may be more important in the development of asthma later in life.

Our interpretation of the Anderson et al. [26,27] meta-analysis comes down on the side that the random p-values indicating no cause−effect associations are more plausible and that their meta-analysis will not likely replicate in the absence of bias. In any case, the linkage between ambient air quality early in life and the development of asthma later in life remains unclear.

One of the statistical methods that we used – estimation of analysis search spaces – indicated that the numbers of statistical tests possible in the base papers were large – median 13,824 (interquartile range 1,536−221,184; range 96−42M) in comparison to actual statistical test results presented. Given such large search spaces, p-hacking cannot be ruled out as an explanation for small p-values reported in the base studies and there is little assurance that the estimates drawn from the base papers into the meta-analysis are unbiased.

The other statistical method that we used – construction of a p-value plot – showed that heterogeneity of the Anderson et al. NO2 results across studies is consistent with a two-component mixture. It makes no sense to average across a mixture. The shape of the p-value plot for NO2 appears consistent with the possibility of analysis manipulation to obtain small p-values in several of the cohort studies. As for PM2.5 results, all corresponding p-values fall on a 45-degree line in the p-value plot indicating complete randomness rather than a true association.

Based on our findings, we conclude that claims made in the base papers used for meta-analysis by Anderson et al. are unreliable due to bias induced by multiple testing and multiple modelling, MTMM. We also show there is evidence that the heterogeneity across base papers used for meta-analysis is more complex than simple sampling from a normal process. The two statistical methods that we used are offered to researchers to independently examine the reliability of other meta-analysis making research claims. We believe that these methods can be used as an aid to judge whether a meta-analysis will replicate.

**Acknowledgements**

The authors gratefully acknowledge Dr. John Dunn, a physician in family medicine, emergency medicine and legal medicine, for helpful discussions on the characteristics of asthma. The authors also gratefully acknowledge comments provided by several reviewers selected by the Editor and anonymous to the authors. The comments helped improve the quality of the article.

**Funding**

No external funding was provided for this study. The study was conceived based on previous work undertaken by CG Stat for the National Association of Scholars (nas.org), New York, NY.

**Conflicts of interest**: The authors have declared no conflict of interest.

Table 3. Counts and analysis search spaces for base papers considered by Anderson et al. [7,8] in their meta-analysis.

| RowID | Study cohort[1] | Outcomes | Predictors | Lags | Covariates | Questions | Models | Search space |
|---|---|---|---|---|---|---|---|---|
| 1 | BAMSE | 7 | 3 | 4 | 6 | 84 | 64 | 5,376 |
| 2 | British Columbia | 1 | 8 | 4 | 7 | 32 | 128 | 4,096 |
| 3 | CHS | 1 | 2 | 8 | 15 | 16 | 32,768 | 524,288 |
| 4 | CHS | 1 | 6 | 5 | 10 | 30 | 1,024 | 30,720 |
| 5 | CHS 2003 | 1 | 5 | 3 | 15 | 15 | 32,768 | 491,520 |
| 6 | CHIBA | 3 | 1 | 3 | 6 | 9 | 64 | 576 |
| 7 | CHIBA | 1 | 3 | 6 | 6 | 18 | 64 | 1,152 |
| 8 | CHIBA | 5 | 4 | 4 | 8 | 80 | 256 | 20,480 |
| 9 | ECHRS | 1 | 1 | 6 | 11 | 6 | 2,048 | 12,288 |
| 10 | GINIplus+LISAplus | 4 | 4 | 6 | 12 | 96 | 4,096 | 393,216 |
| 11 | MISSEB | 1 | 2 | 7 | 6 | 14 | 64 | 896 |
| 12 | OLIN | 1 | 3 | 4 | 5 | 3 | 32 | 96 |
| 13 | OSLO | 4 | 2 | 3 | 11 | 24 | 2,048 | 49,152 |
| 14 | PIAMA | 8 | 4 | 4 | 18 | 128 | 262,144 | 33,000,000 |
| 15 | PIAMA | 5 | 4 | 8 | 18 | 160 | 262,144 | 42,000,000 |
| 16 | RHINE | 1 | 2 | 1 | 8 | 2 | 256 | 512 |
| 17 | TRAPCA | 6 | 3 | 6 | 7 | 108 | 128 | 13,824 |
| 18 | TRAPCA | 7 | 3 | 4 | 9 | 84 | 512 | 43,008 |
| 19 | AHSMOG | 1 | 3 | 3 | 7 | 15 | 128 | 1,920 |

[1]Refer to SI 1 for a listing of the 19 base papers.
Note: Questions = Outcomes x Predictors x Lags. Models = $2^k$ where $k$ = number of Covariates. Search space = approximation of analysis search space = Questions x Models.

Table 4. Summary statistics for counts estimated for 19 base papers considered by Anderson et al. [7,8] in their meta-analysis.

| Statistic | Outcomes | Predictors | Covariates | Lags | Questions | Models | Search space |
|---|---|---|---|---|---|---|---|
| Maximum | 8 | 8 | 8 | 18 | 160 | 262,144 | 42,000,000 |
| Upper quartile | 5 | 4 | 6 | 12 | 84 | 3,072 | 221,184 |
| Median | 1 | 3 | 4 | 8 | 24 | 256 | 13,824 |
| Lower quartile | 1 | 2 | 4 | 7 | 15 | 96 | 1,536 |
| Minimum | 1 | 1 | 1 | 5 | 2 | 32 | 96 |

Note: Questions = Outcomes x Predictors x Lags. Models = $2^k$ where $k$ = number of Covariates. Search space = approximation of analysis search space = Questions x Models.

Table 5. Effect estimate (EE), lower confidence level ($CL_{low}$) and upper confidence level ($CL_{high}$) values and corresponding p-values estimated for cohort studies used by Anderson et al. [7,8] in their meta-analysis.

| Air Component | Study cohort[1] | EE | $CL_{low}$ | $CL_{high}$ | p-value |
|---|---|---|---|---|---|
| NO2 | BAMSE, wheeze | 1.01 | 0.98 | 1.04 | 0.5135 |
| | British Columbia, asthma | 1.13 | 1.04 | 1.23 | 0.0073 |
| | CHS 2003, asthma | 1.03 | 1.01 | 1.05 | 0.0033 |
| | CHS, asthma | 1.24 | 1.06 | 1.46 | 0.0187 |
| | CHIBA, asthma | 1.32 | 1.02 | 1.71 | 0.0691 |
| | ECRHS, asthma | 1.43 | 1.02 | 2.00 | 0.0854 |
| | KRAMER, asthma | 1.19 | 0.85 | 1.68 | 0.3695 |
| | MISSEB, asthma | 1.32 | 0.73 | 2.41 | 0.4553 |
| | OLIN, asthma | 1.00 | 0.35 | 2.87 | 1.0000 |
| | Oslo Birth Cohort, asthma | 0.93 | 0.85 | 1.00 | 0.0673 |
| | PIAMA, asthma | 1.16 | 0.96 | 1.41 | 0.1634 |
| | RHINE, asthma | 1.46 | 1.07 | 1.99 | 0.0500 |
| | TRACPA, asthma | 0.71 | 0.14 | 3.48 | 0.7336 |
| PM2.5 | AHSMOG, asthma | 1.08 | 0.85 | 1.38 | 0.5541 |
| | British Columbia, asthma | 1.10 | 0.90 | 1.35 | 0.3837 |
| | CHIBA, asthma | 1.86 | 0.90 | 3.86 | 0.2547 |
| | PIAMA, asthma | 2.06 | 0.91 | 4.66 | 0.2678 |
| | TRACAP, asthma | 1.60 | 0.45 | 5.70 | 0.6542 |

[1]Study cohorts presented here are in the exact same order as presented in the Anderson et al. [7] Fig. 1 for $NO_2$ study cohorts and Fig. 2 for $PM_{2.5}$ study cohorts.

Table 1. Proposed risk factors for development of asthma in children & adults.

| *1) Induction (sensitization) phase* | *(2) Maintenance (progression) phase* |
| --- | --- |
| Host factors | Environmental stimuli |
| genetic predisposition | allergens |
| gender | respiratory infections (viruses, bacteria) |
| pre-term delivery (e.g., prematurity causes lung problems) | tobacco smoke |
| | indoor/outdoor air quality |
| | occupational sensitizers |
| Environmental stimuli | |
| allergens | |
| respiratory infections (viruses) | |

Note: (1) after Bernstein [32], Scherzer and Grayson [33], Lemanske and Busse [35,36], Singh and Busse [37], Sykes and Johnston [38], Szefler [39], Gelfand [40] Noutsios and Floros [41]. (2) All these factors have support in the scientific literature; some are based on associations and some may be directly causative.

Table 2. Change in US annual national air pollutant concentration averages [44] and annual prevalence of asthma in the US (population-weighted) [42,43] over the 27-year period 1990−2017.

| **Parameter** | **Change** |
|---|---|
| Nitrogen Dioxide (NO2) 1-hour | −50% |
| Nitrogen Dioxide (NO2) annual | −56% |
| Particulate Matter 2.5 microns (PM2.5) 24-hour | −40% |
| Particulate Matter 2.5 microns (PM2.5) annual | −41% |
| Particulate Matter 10 microns (PM10) 24-hour | −34% |
| Carbon Monoxide (CO) 8-hour | −77% |
| Ozone (O3) 8-hour | −22% |
| Sulfur Dioxide (SO2) 1-hour | −88% |
| Prevalence of asthma | +88% |

List of figures:

Figure 1. Proposed childhood asthma causes and risks [30].

Figure 2. Annual prevalence of asthma in the US, population-weighted, 1990−2017 as reported by National Center for Health Statistics, National Health Interview Surveys [42,43].

Figure 3. P-value plot for the Anderson et al. [7,8] meta-analysis (note: solid circles (●) are NOx p-values; open circles (o) are PMx p-values).

Figure 4. Factors to consider when evaluating meta-analysis results presenting as a bi-linear p-value plot.

Fig. 3

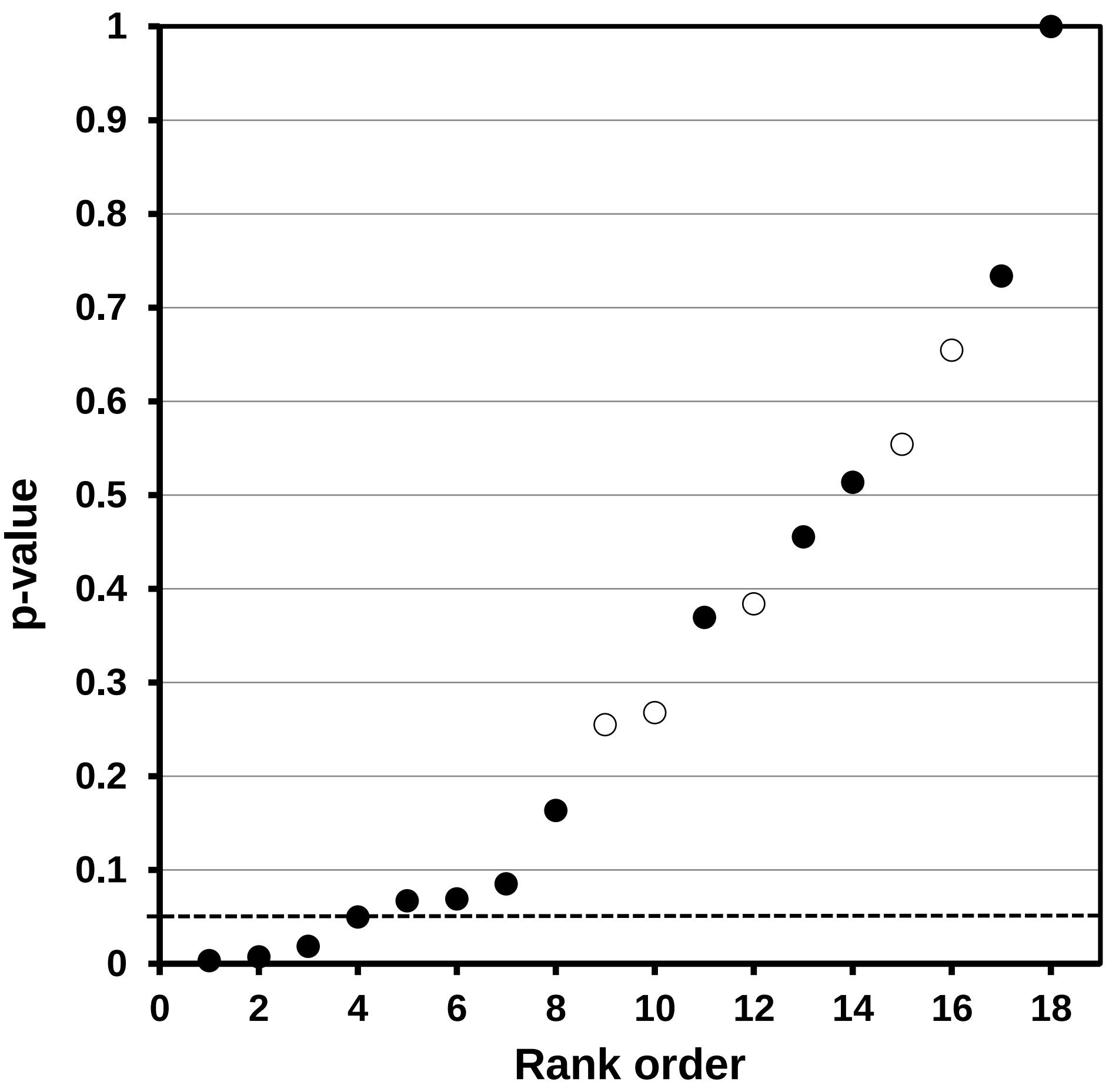

Fig. 1

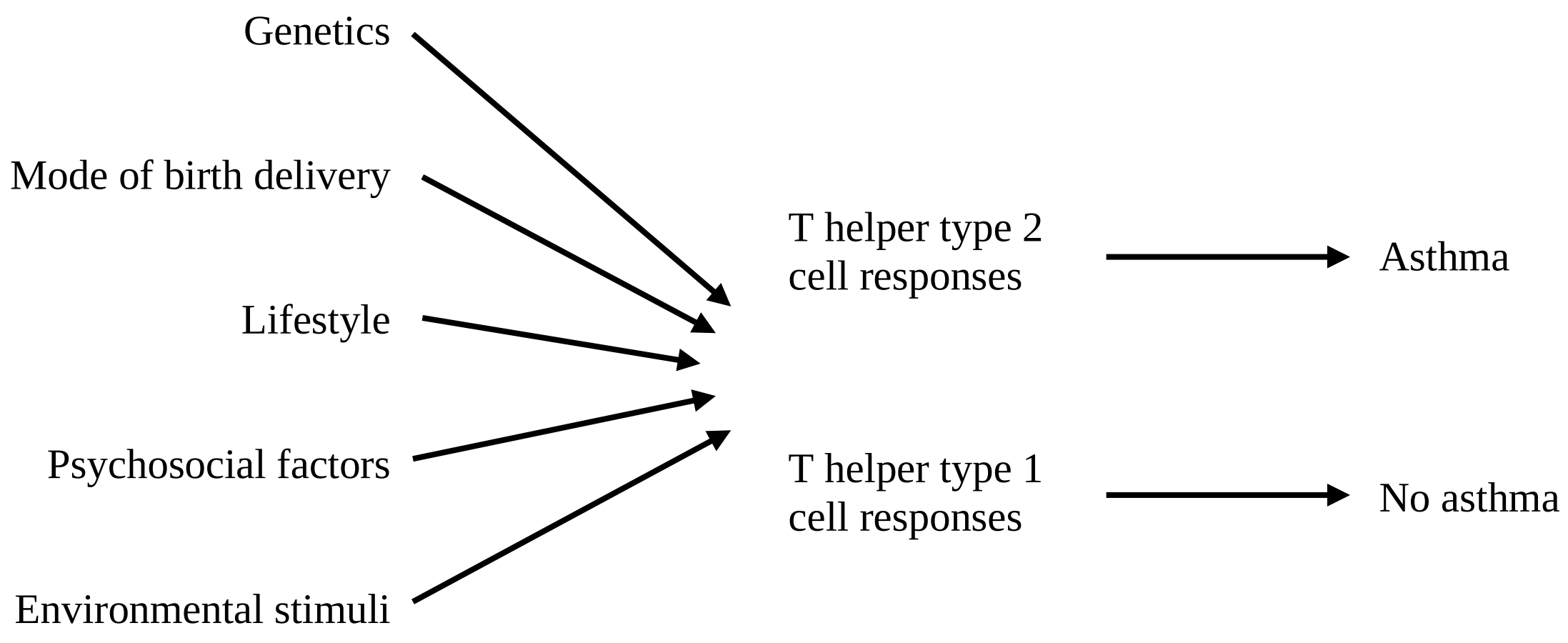

Genetics
Mode of birth delivery
Lifestyle
Psychosocial factors
Environmental stimuli
T helper type 2 cell responses
Asthma
T helper type 1 cell responses
No asthma

Fig 2.

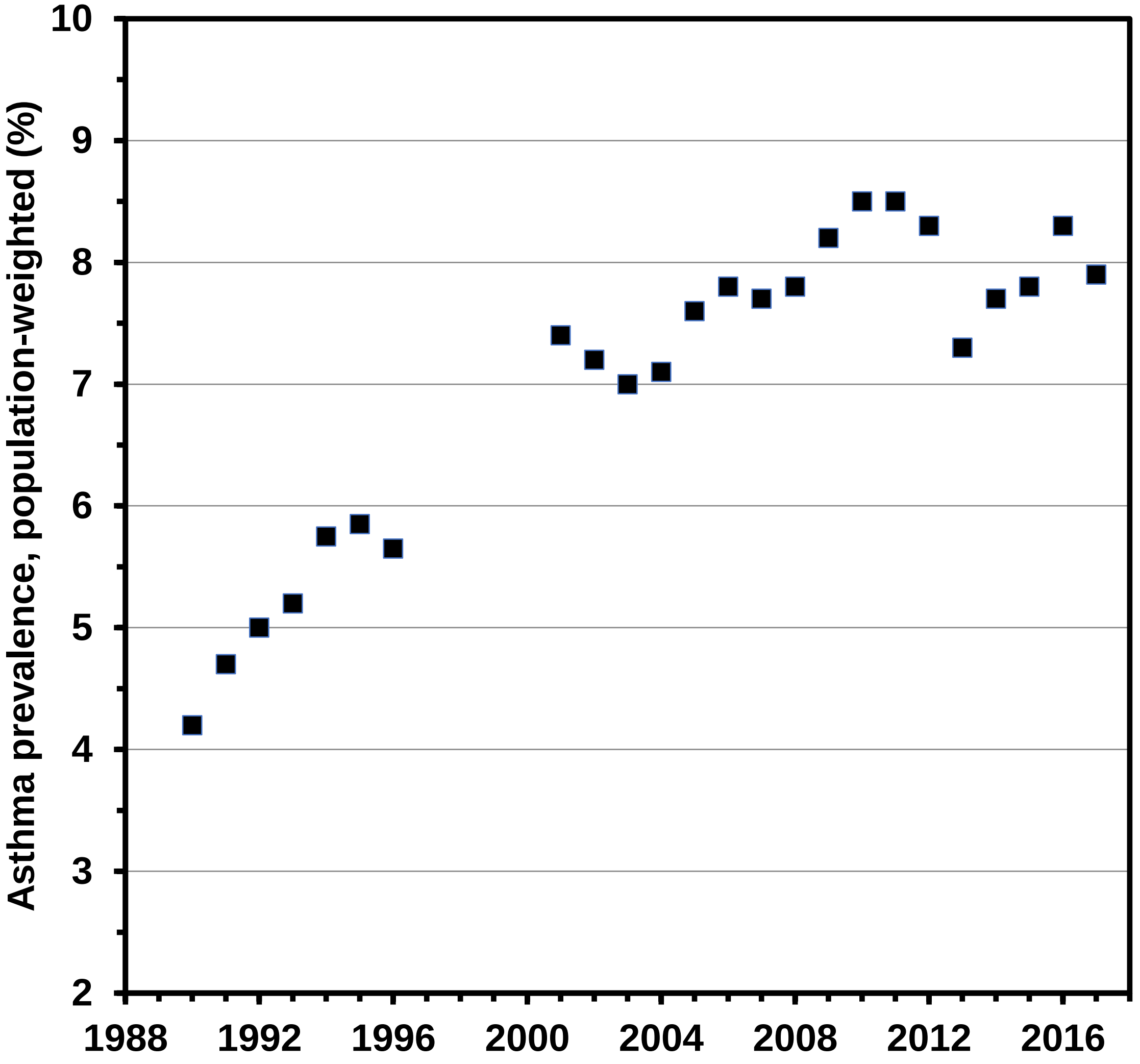

Fig. 4

| Possibility 1 | Possibility 2 |
|---|---|
| **Small p-values true** | **Small p-values false** |
| Most of the studies show a small p-value | Evidence of MTMM (p-hacking) including lags |
| There will be supporting literature | Covariates correlated with outcome, bias |
| There will be a reasonable etiology | Very large sample size elevates small bias to cause |
| No evidence of MTMM (p-hacking) | A number of Bradford Hill criterion not met |
| Most Bradford Hill criterion met | |
| **Large p-values false** | **Large p-values true** |
| Poor research technique | Distribution of large p-values is uniform |
| Underpowered studies | Good negative effect studies |
| Masking covariates hide real effect | No clear etiology |
| Role of chance | |

# Reliability of meta-analysis of an association between ambient air quality and development of asthma later in life

## Supplementary Information

S. Stanley Young,[1] Kai-Chieh Cheng,[2] Jin Hua Chen,[2] Shu-Chuan Chen,[3] Warren B. Kindzierski[5]

**Affiliations**

[1]CGStat, Raleigh, NC, USA

[2]Graduate Institute of Data Science, Taipei Medical University, Taipei City, Taiwan

[3]Department of Mathematics and Statistics, Idaho State University, Pocatello, ID, USA

[4]School of Public Health, University of Alberta, Edmonton, Alberta, Canada

**Corresponding author:**

Warren B. Kindzierski, School of Public Health, University of Alberta, Edmonton, Alberta, T6G 1C9, Canada. Email: warrenk@ualberta.ca.; phone: 780-492-0382; fax: 780-492-0364.

**16 pages**

Background information on meta-analysis evaluation is freely available at https://arxiv.org/abs/1808.04408.

## Supplemental Information 1

## Base Papers used for Search Space Counting

## Supplemental Information 2

## Three examples of search space analysis of hypothetical observational (cohort) studies of ambient air quality and development of asthma

---

Questions = Number of questions at issue = outcomes x predictors x lags
Models = Number of questions at issue = $2^k$ where $k$ = number of covariates
Search space = Approximation of analysis search space (number of statistical tests) = Questions x Models

---

In the three examples below, assume the following conditions, used differently in each study:

- prospective cohort studies of air pollution and respiratory health with sufficiently large sample sizes of children enrolled in elementary grade 4 and followed for 8 years
- one outcome is of interest – asthma diagnosis later in life (surveyed annually during grade 5 until high school graduation) due to inferred exposure to outdoor air quality predictors
- three outdoor air quality predictors – PM2.5, NO2 and O3
- an 'asthma diagnosis' is based on responses to annual questionnaires where a subject in the cohort answers yes to the following question… "*Has a doctor ever said you had asthma?*"
- questionnaires were administered by trained professionals once each year to each family of an enlisted child for up to 8 years of follow-up until high school graduation (i.e., 8 lags)
- covariate confounders were considered related to four socioeconomic measures – median household income, proportion of subjects with no high school diploma, percent of males unemployed, percent living in poverty
- covariate confounders were considered related to two meteorological variables – humidity and temperature
- air quality parameters were also be treated as covariate confounders

---

***Example 1***

Assume a bare bones epidemiological investigation of childhood asthma diagnosis using the 3 air quality predictors – daily average levels of PM2.5, NO2 and O3, 8 years of follow-up and no covariate confounders:

- Questions = 1 outcome x 3 predictors x 8 lags = 24
- Models = 1 (i.e., no consideration of covariate confounding)
- Search space = approximation of analysis search space = 24 x 1 = 24

---

***Example 2***

Assume an epidemiological investigation of the same 3 predictors, 8 years of follow-up (i.e., lags) and 2 weather variables treated as covariate confounders (annual average temperature and annual average humidity) and 4 socioeconomic measures also treated as covariate confounders:

- Questions = 1 outcome x 3 predictors x 8 lags = 24
- Models = $2^{2+4}$ = 64
- Search space = approximation of analysis search space = 24 x 64 = 1,536

---

***Example 3***

Assume an epidemiological analysis of the same 3 predictors, 8 years of follow-up (i.e., lags) now using 4 weather variables treated as covariate confounders (average temperature in winter and summer months, average humidity in winter and summer months), plus the 4 socioeconomic measures treated as covariate confounders, and using air quality parameters also treated as covariate confounders:

- Questions = 1 outcome x 3 predictors x 8 lags = 24
- Models* = $2^{4+4+2}$ = 1,024
- Search space = approximation of analysis search space = 24 x 1,024 = 24,576

* Note: there are 10 covariates in Models – 4 weather variables, 4 socioeconomic measures; with 1 air quality variable is treated as a predictor which is adjusted with the other 2 air quality variables as covariates

---

## Supplemental Information 3

## Summary Statistics of Datasets from Meta-analysis of Selected Cancers in Petroleum Refinery Workers after Schnatter et al. [53]

Note: Base study=base study 1st author name in Schnatter et al. [53]; RR=relative risk; LCL=lower confidence limit; UCL=upper confidence limit; bold, italicized p-value <0.05; p-values were calculated using the method of Altman DG, Bland JM. How to obtain the p-value from a confidence interval. *British Medical Journal*. 2011;343: d2304. doi:10.1136/bmj.d2304. A p-value calculated as ≤0.0001 was recorded as 0.0001.

Chronic myeloid leukemia risk for petroleum refinery workers:

| **Base study** | **RR** | **LCL** | **UCL** | **p-value** |
|---|---|---|---|---|
| Collingwood 1996 | 0.53 | 0.07 | 3.74 | 0.54263 |
| Divine 1999a | 1.05 | 0.60 | 1.85 | 0.87478 |
| Gun 2006b | 1.09 | 0.45 | 2.61 | 0.85800 |
| Huebner 2004 | 1.68 | 0.88 | 3.23 | 0.11778 |
| Lewis 2000a | 1.08 | 0.35 | 3.35 | 0.90196 |
| Rushton 1993a | 0.89 | 0.50 | 1.61 | 0.70923 |
| Satin 1996 | 0.85 | 0.38 | 1.88 | 0.70346 |
| Satin 2002 | 0.45 | 0.14 | 1.39 | 0.17355 |
| Tsai 2007 | 0.66 | 0.21 | 2.05 | 0.48425 |
| Wong 2001a | 1.31 | 0.55 | 3.15 | 0.55549 |
| Wong 2001b | 1.96 | 0.49 | 7.84 | 0.34689 |
| Wongsrichanalia 1989 | 0.44 | 0.06 | 3.12 | 0.42318 |

Mesothelioma risk for petroleum refinery workers (based on mesothelioma subgroup analysis using Schnatter et al. [53] 'Best Methods' dataset):

| **Base study** | **RR** | **LCL** | **UCL** | **p-value** |
|---|---|---|---|---|
| Devine 1999a | 2.97 | 2.21 | 3.99 | ***0.0001*** |
| Gamble 2000 | 2.43 | 1.35 | 4.39 | ***0.00321*** |
| Gun 2006a | 3.77 | 2.14 | 6.64 | ***0.0001*** |
| Honda 1995 | 2.00 | 1.04 | 3.84 | ***0.03720*** |
| Hornstra 1993 | 5.51 | 3.38 | 8.99 | ***0.0001*** |
| Huebner 2009 | 2.44 | 1.83 | 3.24 | ***0.0001*** |
| Kaplan 1986 | 2.41 | 1.26 | 4.64 | ***0.00817*** |
| Lewis 2000a | 8.68 | 5.77 | 13.06 | ***0.0001*** |
| Tsai 2003 | 2.16 | 0.70 | 6.69 | 0.18215 |
| Tsai 2007 | 2.50 | 1.63 | 3.83 | ***0.0001*** |

# Summary Statistics of Datasets from Meta-analysis of Elderly Long-term Exercise Training−Mortality & Morbidity Risk after de Souto Barreto et al. [54]

Note: Base study=base study 1st author name in de Souto Barreto et al. [54]; RR=relative risk; LCL=lower confidence limit; UCL=upper confidence limit; bold, italicized p-value <0.05; p-values were calculated using the method of Altman DG, Bland JM. How to obtain the p-value from a confidence interval. *British Medical Journal*. 2011;343: d2304. doi:10.1136/bmj.d2304. A p-value calculated as ≤0.0001 was recorded as 0.0001.

| **Outcome** | **No.** | **Base Study ID (n=69)** | **RR** | **LCL** | **UCL** | ***p-value*** |
|---|---|---|---|---|---|---|
| Mortality | 1 | Belardinelli et al. 2012 | 0.38 | 0.13 | 1.15 | 0.08153 |
| | 2 | Barnett et al. 2003 | 0.14 | 0.01 | 2.63 | 0.16736 |
| | 3 | O'Connor et al. 2009 | 0.96 | 0.80 | 1.16 | 0.67980 |
| | 4 | Campbell et al. 1997 | 0.50 | 0.09 | 2.70 | 0.43245 |
| | 5 | El−Khoury et al. 2015 | 0.84 | 0.26 | 2.72 | 0.78350 |
| | 6 | Galvão et al. 2014 | 3.00 | 0.13 | 71.92 | 0.50544 |
| | 7 | Gianoudis et al. 2014 | 1.00 | 0.06 | 15.72 | 1.00000 |
| | 8 | Hewitt et al. 2018 | 1.02 | 0.52 | 2.03 | 0.95866 |
| | 9 | Karinkanta et al. 2007 | 0.33 | 0.01 | 7.93 | 0.52568 |
| | 10 | Kemmler et al. 2010 | 0.33 | 0.01 | 8.10 | 0.52704 |
| | 11 | King et al. 2002 | 0.32 | 0.01 | 7.68 | 0.51168 |
| | 12 | Kovács et al. 2013 | 0.40 | 0.14 | 1.17 | 0.09039 |
| | 13 | Lam et al. 2012 | 0.64 | 0.06 | 7.05 | 0.72673 |
| | 14 | Lam et al. 2015 | 0.30 | 0.03 | 2.82 | 0.30309 |
| | 15 | Lord et al. 2003 | 4.84 | 0.55 | 42.33 | 0.15519 |
| | 16 | Merom et al. 2015 | 1.36 | 0.22 | 8.23 | 0.75225 |
| | 17 | Pahor et al. 2006 | 0.99 | 0.14 | 6.97 | 0.99276 |
| | 18 | Pahor et al. 2014/Gill et al. 2016 | 1.14 | 0.76 | 1.71 | 0.53739 |
| | 19 | Patil et al. 2015 | 0.11 | 0.01 | 2.04 | 0.10355 |
| | 20 | Pitkälä et al. 2013 | 0.25 | 0.06 | 1.14 | 0.06455 |
| | 21 | Prescott et al. 2008 | 0.42 | 0.08 | 2.12 | 0.30363 |
| | 22 | Rejeski et al. 2017 | 0.34 | 0.01 | 8.16 | 0.53914 |
| | 23 | Rolland et al. 2007 | 0.88 | 0.34 | 2.28 | 0.80441 |
| | 24 | Sherrington et al. 2014 | 1.10 | 0.46 | 2.63 | 0.84135 |
| | 25 | Underwood et al. 2013 | 1.06 | 0.84 | 1.35 | 0.64301 |
| | 26 | Van Uffelen et al. 2008 | 0.36 | 0.01 | 8.72 | 0.56573 |
| | 27 | von Stengel et al. 2011 | 0.34 | 0.01 | 8.15 | 0.53907 |
| | 28 | Voukelatos et al. 2015 | 9.09 | 0.49 | 167.75 | 0.13843 |
| | 29 | Wolf et al. 2003 | 0.97 | 0.14 | 6.86 | 0.97786 |

**(continued)**

| Outcome | No. | Base Study ID (n=69) | RR | LCL | UCL | *p-value* |
|---|---|---|---|---|---|---|
| Hospitalization | 30 | Belardinelli et al. 2012 | 0.30 | 0.15 | 0.62 | ***0.00092*** |
| | 31 | O’Connor et al. 2009 | 0.97 | 0.91 | 1.03 | 0.34039 |
| | 32 | Hambrecht et al. 2004 | 0.16 | 0.02 | 1.31 | 0.08551 |
| | 33 | Hewitt et al. 2018 | 0.64 | 0.27 | 1.50 | 0.31209 |
| | 34 | Kovács et al. 2013 | 2.00 | 0.19 | 21.21 | 0.57619 |
| | 35 | Messier et al. 2013 | 8.54 | 0.46 | 157.06 | 0.14992 |
| | 36 | Mustata et al. 2011 | 0.33 | 0.02 | 7.32 | 0.47075 |
| | 37 | Pahor et al. 2006 | 0.99 | 0.68 | 1.44 | 0.96195 |
| | 38 | Pahor et al. 2014/Gill et al. 2016 | 1.10 | 0.99 | 1.22 | 0.07332 |
| | 39 | Pitkala et al. 2013 | 0.78 | 0.55 | 1.12 | 0.17166 |
| | 40 | Rejeski et al. 2017 | 3.04 | 0.13 | 73.46 | 0.50161 |
| | 41 | Rolland et al. 2007 | 1.82 | 0.95 | 3.49 | 0.07083 |
| Injurious falls | 42 | Barnett et al. 2003 | 0.77 | 0.48 | 1.21 | 0.27108 |
| | 43 | Campbell et al. 1997 | 0.67 | 0.45 | 1.00 | ***0.04892*** |
| | 44 | El−Khoury et al. 2015 | 0.90 | 0.78 | 1.05 | 0.16541 |
| | 45 | Hewitt et al. 2018 | 0.58 | 0.42 | 0.81 | ***0.00120*** |
| | 46 | MacRae et al. 1994 | 0.16 | 0.01 | 2.92 | 0.20731 |
| | 47 | Pahor et al. 2014/Gill et al. 2016 | 0.89 | 0.66 | 1.20 | 0.45350 |
| | 48 | Patil et al. 2015 | 0.51 | 0.31 | 0.84 | ***0.00810*** |
| | 49 | Pitkälä et al. 2013 | 0.65 | 0.39 | 1.09 | 0.10016 |
| | 50 | Reinsch et al. 1992 | 1.46 | 0.37 | 5.81 | 0.60232 |
| Fractures | 51 | Belardinelli et al. 2012 | 0.19 | 0.01 | 3.89 | 0.27847 |
| | 52 | O’Connor et al. 2009 | 0.60 | 0.32 | 1.11 | 0.10725 |
| | 53 | El−Khoury et al. 2015 | 0.88 | 0.60 | 1.25 | 0.50488 |
| | 54 | Gianoudis et al. 2014 | 3.00 | 0.12 | 72.57 | 0.51161 |
| | 55 | Hewitt et al. 2018 | 0.80 | 0.20 | 3.11 | 0.76275 |
| | 56 | Karinkanta et al. 2007 | 1.00 | 0.15 | 6.73 | 1.00000 |
| | 57 | Kemmler et al. 2010 | 0.49 | 0.19 | 1.25 | 0.13795 |
| | 58 | Kovács et al. 2013 | 3.00 | 0.13 | 71.56 | 0.50509 |
| | 59 | Lam et al. 2012 | 1.27 | 0.06 | 28.95 | 0.88844 |
| | 60 | Pahor et al. 2014/Gil et al. 2016 | 0.87 | 0.63 | 1.19 | 0.39774 |
| | 61 | Patil et al. 2015 | 0.66 | 0.28 | 1.59 | 0.35403 |
| | 62 | Pitkälä et al. 2013 | 1.00 | 0.26 | 3.84 | 1.00000 |
| | 63 | Reinsch et al. 1992 | 0.45 | 0.04 | 4.78 | 0.52344 |
| | 64 | Rolland et al. 2007 | 2.50 | 0.50 | 12.44 | 0.26692 |
| | 65 | Sherrington et al. 2014 | 0.92 | 0.46 | 1.85 | 0.82585 |
| | 66 | Underwood et al. 2013 | 1.05 | 0.63 | 1.74 | 0.86094 |
| | 67 | Villareal et al. 2011 | 0.52 | 0.05 | 5.39 | 0.59601 |
| | 68 | von Stengel et al. 2011 | 0.58 | 0.18 | 1.87 | 0.36778 |
| | 69 | Wolf et al. 2003 | 0.78 | 0.17 | 3.67 | 0.76405 |

## Summary Statistics of Datasets from Meta-analysis of

## Smoking−Lung Squamous Cell Carcinoma Risk after Lee et al. [55]

Note: Base study=base study 1st author name in Lee et al. [55]; RR=relative risk; LCL=lower confidence limit; UCL=upper confidence limit; p-value calculated after Altman (2011); bold, italicized p-value <0.05; p-values were calculated using the method of Altman DG, Bland JM. How to obtain the p-value from a confidence interval. *British Medical Journal*. 2011;343: d2304. doi:10.1136/bmj.d2304. A p-value calculated as ≤0.0001 was recorded as 0.0001.

| Place | No. | Base Study ID (n=102) | RR | LCL | UCL | *p-value* |
|---|---|---|---|---|---|---|
| USA | 1 | 1948 WYNDE4 m | 12.79 | 6.19 | 26.14 | ***0.0001*** |
| | 2 | 1948 WYNDE4 f | 2.82 | 2.55 | 13.31 | ***0.01380*** |
| | 3 | 1949 BRESLO c | 3.69 | 2.06 | 6.62 | ***0.0001*** |
| | 4 | 1952 HAMMON m | 16.88 | 6.29 | 45.29 | ***0.0001*** |
| | 5 | 1955 HAENSZ f | 3.00 | 1.90 | 4.73 | ***0.0001*** |
| | 6 | 1957 BYERS1 m | 8.29 | 5.29 | 13.00 | ***0.0001*** |
| | 7 | 1960 LOMBA2 f | 4.24 | 2.40 | 7.50 | ***0.0001*** |
| | 8 | 1962 WYNDE2 m | 19.72 | 6.21 | 62.59 | ***0.0001*** |
| | 9 | 1964 OSANN2 f | 35.10 | 4.80 | 256.00 | ***0.00048*** |
| | 10 | 1966 WYNDE3 m | 18.29 | 5.71 | 58.56 | ***0.0001*** |
| | 11 | 1966 WYNDE3 f | 6.79 | 2.45 | 18.82 | ***0.00025*** |
| | 12 | 1968 HINDS f | 16.13 | 7.66 | 33.97 | ***0.0001*** |
| | 13 | 1969 STAYNE m | 3.47 | 2.17 | 5.56 | ***0.0001*** |
| | 14 | 1969 WYNDE6 m | 18.59 | 12.74 | 27.13 | ***0.0001*** |
| | 15 | 1969 WYNDE6 f | 32.37 | 17.66 | 59.35 | ***0.0001*** |
| | 16 | 1975 COMSTO m | 8.07 | 1.91 | 34.02 | ***0.00452*** |
| | 17 | 1975 COMSTO f | 46.20 | 2.74 | 778.83 | ***0.00784*** |
| | 18 | 1976 BUFFLE m | 14.03 | 4.73 | 41.61 | ***0.0001*** |
| | 19 | 1976 BUFFLE f | 13.04 | 3.99 | 42.66 | ***0.0001*** |
| | 20 | 1979 CORREA c | 28.30 | 18.60 | 43.20 | ***0.0001*** |
| | 21 | 1979 SIEMIA m | 22.70 | 6.90 | 75.20 | ***0.0001*** |
| | 22 | 1980 DORGAN m | 18.90 | 7.00 | 51.30 | ***0.0001*** |
| | 23 | 1980 DORGAN f | 11.10 | 7.20 | 17.10 | ***0.0001*** |
| | 24 | 1981 JAIN m | 18.00 | 5.50 | 111.00 | ***0.00018*** |
| | 25 | 1981 JAIN f | 25.50 | 7.93 | 156.00 | ***0.0001*** |
| | 26 | 1981 WU f | 24.29 | 3.40 | 173.76 | ***0.00153*** |
| | 27 | 1983 BAND m | 37.45 | 17.60 | 79.58 | ***0.0001*** |
| | 28 | 1984 BROWN2 m | 11.10 | 9.50 | 12.90 | ***0.0001*** |
| | 29 | 1984 BROWN2 f | 20.10 | 16.40 | 24.80 | ***0.0001*** |

**(continued)**

| Place | No. | Base Study ID (n=102) | RR | LCL | UCL | *p-value* |
|---|---|---|---|---|---|---|
| USA | 30 | 1984 OSANN m | 36.10 | 17.80 | 73.30 | ***0.0001*** |
| | 31 | 1984 OSANN f | 26.40 | 14.50 | 48.10 | ***0.0001*** |
| | 32 | 1984 SCHWAR m1 | 32.81 | 4.48 | 240.23 | ***0.0001*** |
| | 33 | 1984 SCHWAR m2 | 1.81 | 0.50 | 6.78 | 0.37881 |
| | 34 | 1984 SCHWAR f1 | 43.23 | 2.60 | 718.15 | ***0.00862*** |
| | 35 | 1984 SCHWAR f2 | 62.61 | 3.64 | 1076.10 | ***0.00441*** |
| | 36 | 1985 KHUDER m | 7.82 | 3.87 | 15.77 | ***0.0001*** |
| | 37 | 1986 ANDERS f | 25.57 | 10.29 | 63.56 | ***0.0001*** |
| | 38 | 1989 HEGMAN c | 30.80 | 12.48 | 76.03 | ***0.0001*** |
| Europe | 39 | 1947 ORMOS m | 10.14 | 2.41 | 42.79 | ***0.00165*** |
| | 40 | 1948 DOLL m | 13.17 | 4.12 | 42.10 | ***0.0001*** |
| | 41 | 1948 DOLL f | 2.13 | 1.06 | 4.27 | ***0.03311*** |
| | 42 | 1948 KREYBE m | 10.87 | 3.47 | 34.04 | ***0.0001*** |
| | 43 | 1948 KREYBE f | 2.29 | 0.89 | 5.88 | 0.08506 |
| | 44 | 1954 STASZE m | 57.77 | 3.58 | 933.17 | ***0.00430*** |
| | 45 | 1954 STASZE f | 32.45 | 1.32 | 800.04 | ***0.03297*** |
| | 46 | 1959 TIZZAN c | 2.70 | 1.99 | 3.67 | ***0.0001*** |
| | 47 | 1964 ENGELA m | 6.45 | 1.97 | 21.11 | ***0.00211*** |
| | 48 | 1966 TOKARS c | 6.80 | 1.20 | 38.70 | ***0.03026*** |
| | 49 | 1971 NOU m | 27.17 | 6.60 | 11.85 | ***0.0001*** |
| | 50 | 1971 NOU f | 7.09 | 1.35 | 37.19 | ***0.02043*** |
| | 51 | 1972 DAMBER m | 11.80 | 6.40 | 23.00 | ***0.0001*** |
| | 52 | 1975 ABRAHA m | 92.66 | 5.77 | 1488.21 | ***0.00143*** |
| | 53 | 1975 ABRAHA f | 5.35 | 2.22 | 12.90 | ***0.00021*** |
| | 54 | 1976 LUBIN2 m | 16.66 | 12.69 | 21.86 | ***0.0001*** |
| | 55 | 1976 LUBIN2 f | 5.78 | 4.34 | 7.71 | ***0.0001*** |
| | 56 | 1977 ALDERS m | 14.70 | 3.40 | 63.64 | ***0.00035*** |
| | 57 | 1977 ALDERS f | 6.09 | 2.68 | 13.82 | ***0.0001*** |
| | 58 | 1979 BARBON m | 14.52 | 6.35 | 33.20 | ***0.0001*** |
| | 59 | 1979 DOSEME m | 3.60 | 2.60 | 5.00 | ***0.0001*** |
| | 60 | 1980 JEDRYC m | 12.84 | 5.58 | 29.55 | ***0.0001*** |
| | 61 | 1983 SVENSS f | 12.62 | 3.97 | 40.14 | ***0.0001*** |
| | 62 | 1985 BECHER f | 10.69 | 2.43 | 47.00 | ***0.00177*** |
| | 63 | 1987 KATSOU f | 6.11 | 2.69 | 13.87 | ***0.0001*** |
| | 64 | 1988 JAHN m | 23.03 | 7.29 | 72.81 | ***0.0001*** |
| Asia | 65 | 1961 ISHIMA c | 21.00 | 3.38 | 868.40 | ***0.03122*** |
| | 66 | 1964 JUSSAW m | 25.43 | 13.87 | 46.63 | ***0.0001*** |
| | 67 | 1965 MATSUD m | 39.01 | 5.44 | 279.84 | ***0.00029*** |
| | 68 | 1976 CHAN m | 15.22 | 3.61 | 64.12 | ***0.00023*** |
| | 69 | 1976 CHAN f | 6.44 | 3.44 | 12.06 | ***0.0001*** |
| | 70 | 1976 LAMWK2 m | 6.89 | 2.65 | 17.90 | ***0.0001*** |
| | 71 | 1976 LAMWK2 f | 6.49 | 3.27 | 12.88 | ***0.0001*** |

**(continued)**

| Place | No. | Base Study ID (n=102) | RR | LCL | UCL | *p-value* |
|---|---|---|---|---|---|---|
| Asia | 72 | 1976 TSUGAN m | 14.55 | 0.75 | 283.37 | 0.07657 |
| | 73 | 1978 ZHOU m | 3.14 | 1.90 | 5.18 | ***0.0001*** |
| | 74 | 1978 ZHOU f | 3.81 | 1.50 | 9.68 | ***0.00496*** |
| | 75 | 1981 KOO f | 4.15 | 2.46 | 6.98 | ***0.0001*** |
| | 76 | 1981 LAMWK f | 10.54 | 4.19 | 26.52 | ***0.0001*** |
| | 77 | 1981 XU3 m | 5.90 | 1.69 | 20.57 | ***0.00540*** |
| | 78 | 1981 XU3 f | 25.67 | 4.99 | 131.94 | ***0.00012*** |
| | 79 | 1982 ZHENG m | 16.82 | 6.05 | 46.71 | ***0.0001*** |
| | 80 | 1982 ZHENG f | 5.45 | 3.11 | 9.54 | ***0.0001*** |
| | 81 | 1983 LAMTH f | 8.10 | 4.16 | 15.77 | ***0.0001*** |
| | 82 | 1984 GAO m | 8.40 | 4.70 | 15.00 | ***0.0001*** |
| | 83 | 1984 GAO f | 7.20 | 4.60 | 11.10 | ***0.0001*** |
| | 84 | 1984 LUBIN m | 6.33 | 2.29 | 17.45 | ***0.00040*** |
| | 85 | 1985 CHOI m | 5.45 | 2.34 | 12.67 | ***0.0001*** |
| | 86 | 1985 CHOI f | 6.94 | 2.68 | 17.96 | ***0.0001*** |
| | 87 | 1985 WUWILL f | 4.20 | 3.00 | 5.90 | ***0.0001*** |
| | 88 | 1986 SOBUE m | 17.88 | 7.82 | 40.87 | ***0.0001*** |
| | 89 | 1986 SOBUE f | 8.74 | 5.09 | 15.02 | ***0.0001*** |
| | 90 | 1988 WAKAI m | 8.61 | 2.08 | 35.72 | ***0.00305*** |
| | 91 | 1988 WAKAI f | 25.23 | 6.87 | 92.66 | ***0.0001*** |
| | 92 | 1990 FAN c | 11.68 | 5.04 | 27.04 | ***0.0001*** |
| | 93 | 1990 GER c | 3.19 | 1.08 | 9.42 | ***0.03547*** |
| | 94 | 1990 LUO c | 10.90 | 2.50 | 47.90 | ***0.00157*** |
| | 95 | 1991 KIHARA c | 26.97 | 10.84 | 67.08 | ***0.0001*** |
| | 96 | 1997 SEOW f | 17.50 | 6.95 | 44.09 | ***0.0001*** |
| Other | 97 | 1978 JOLY m | 31.21 | 7.69 | 126.68 | ***0.0001*** |
| | 98 | 1978 JOLY f | 18.56 | 7.74 | 44.51 | ***0.0001*** |
| | 99 | 1987 PEZZOT m | 62.74 | 3.86 | 1019.50 | ***0.00367*** |
| | 100 | 1991 SUZUK2 c | 31.00 | 4.20 | 227.00 | ***0.00078*** |
| | 101 | 1993 DESTE2 m | 13.20 | 4.70 | 37.10 | ***0.0001*** |
| | 102 | 1994 MATOS m | 8.08 | 2.58 | 25.50 | ***0.00038*** |

**Figure SI3−1.** p-Value plots for meta-analysis of small observational datasets representing: (i) plausible true null hypothesis for a petroleum refinery worker–chronic myeloid leukemia causal relationship (n=12) after Schnatter et al. [53] and (ii) plausible true alternative hypothesis for a petroleum refinery worker–mesothelioma causal relationship (n=10) after Schnatter et al. [53].

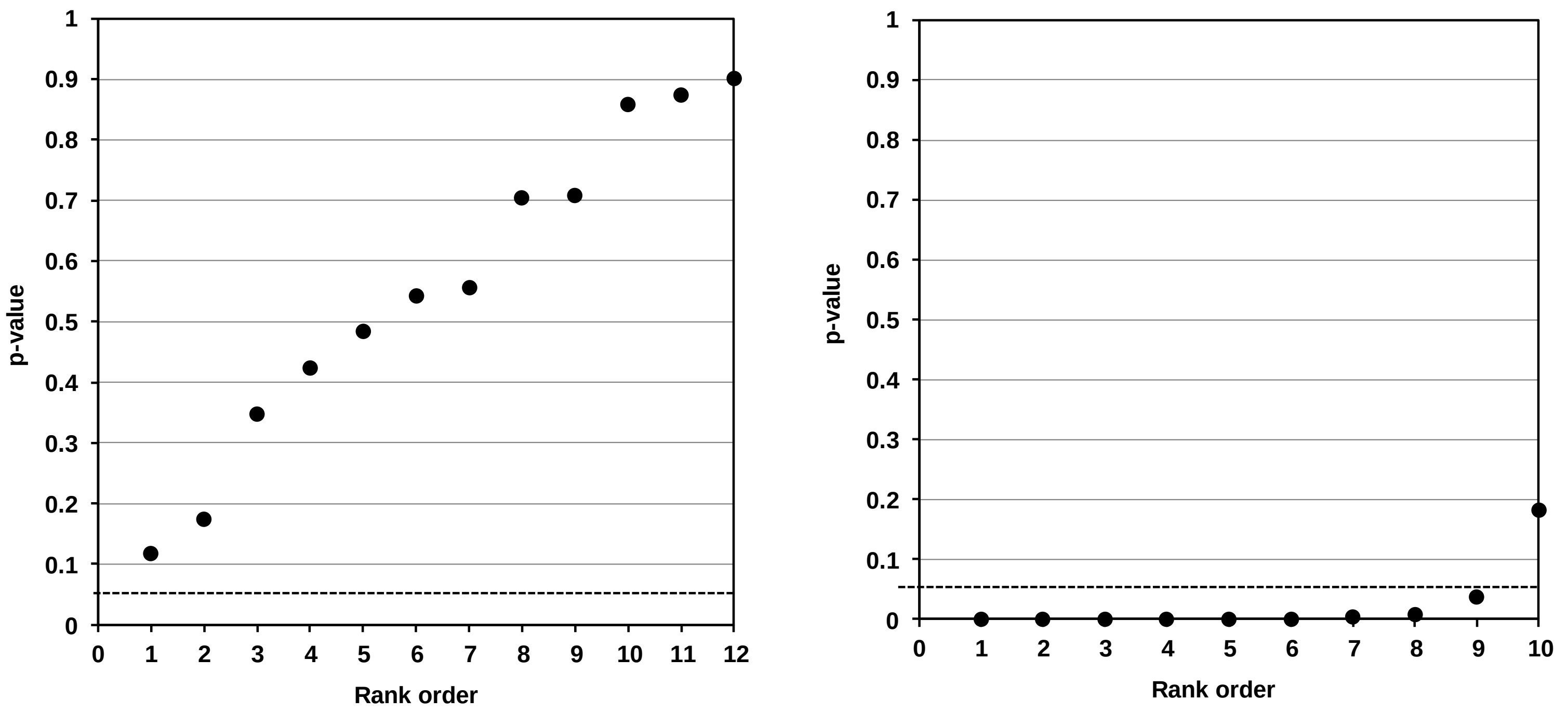

(i) (ii)

**Figure SI3−2.** p-Value plots for meta-analysis of large observational datasets representing: (i) plausible true null hypothesis for an elderly long-term exercise training–mortality & morbidity causal relationship (n=69) after de Souto Barreto et al. [54] and (ii) plausible true alternative hypothesis for a smoking–lung squamous cell carcinoma causal relationship (n=102) after Lee et al. [55].

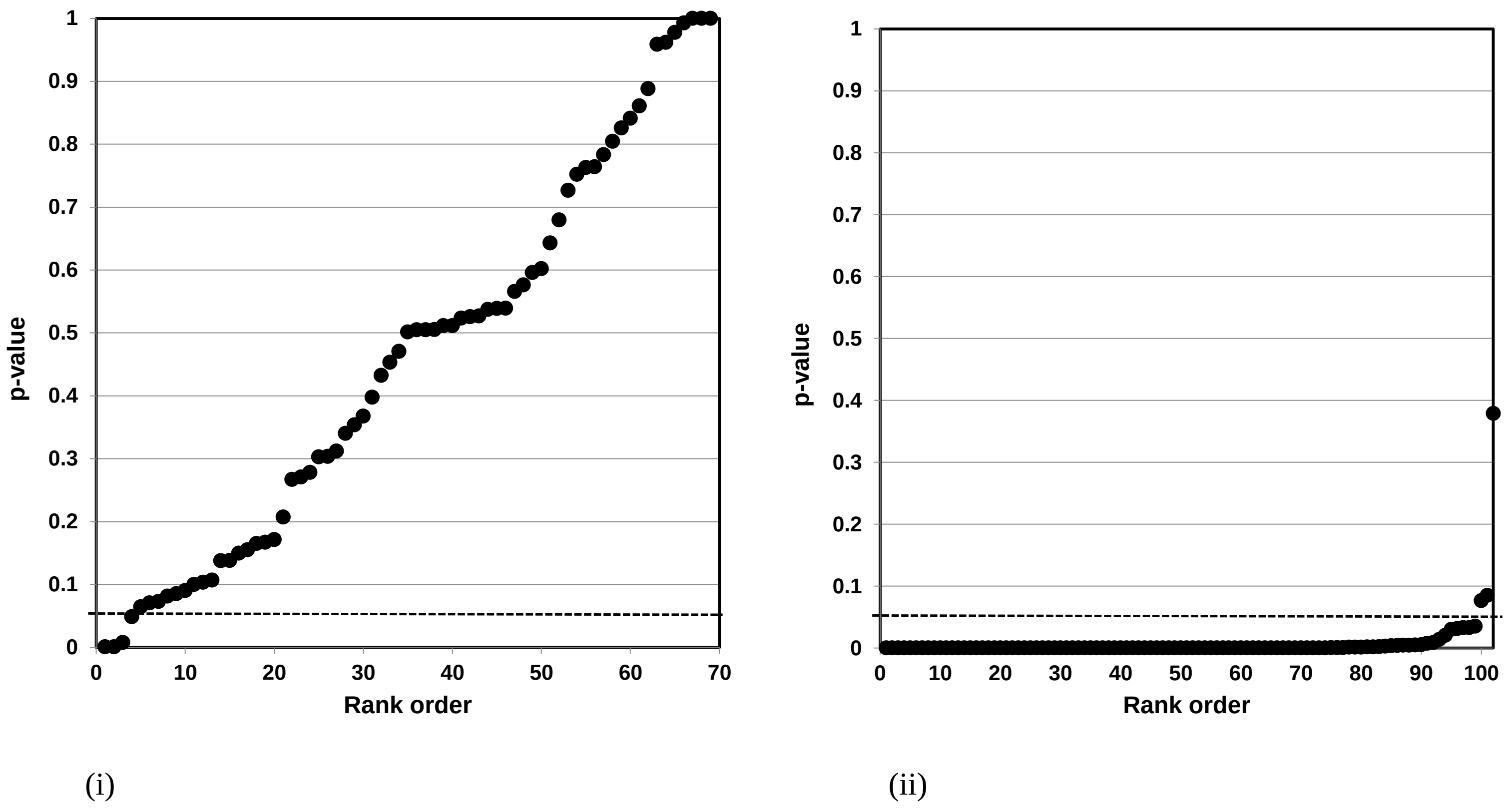

**Explanation of Figures SI3−1 and SI3−3**

Figure SI3−1 presents p-value plots for small meta-analysis datasets (i.e., n<15 base papers) representing selected cancers in petroleum refinery workers after Schnatter et al. [53]:

- Figure 1 left image – presents p-values as a sloped line from left to right at approximately 45-degrees representing a plausible true null hypothesis for a chronic myeloid leukemia causal relationship in petroleum refinery workers (n=12) after Schnatter et al. [53].
- Figure 1 right image – presents a majority of p-values below the .05 line representing a plausible true alternative hypothesis for a mesothelioma causal relationship in petroleum refinery workers (n=10) after Schnatter et al. [53].

Figure SI3−2 presents another set of p-value plots for large meta-analysis datasets (i.e., n>65 base papers) showing plausible true null and true alternative hypothesis:

- Figure 2 left image – presents p-values as a sloped line from left to right at approximately 45-degrees representing a plausible true null hypothesis for an elderly long-term exercise training−mortality & morbidity causal relationship (n=69) after de Souto Barreto et al. [54].
- Figure 2 right image – presents a majority of p-values below the .05 line representing a plausible true alternative hypothesis for a smoking−squamous cell carcinoma causal relationship (n=102) after Lee et al. [55].